\documentclass[epsfig,floats,twocolumn,amssymb,amsmath,floatfix,showpacs]{revtex4}
\usepackage{graphicx}

\begin{document}

\title{Three-Sphere Low Reynolds Number Swimmer with a Cargo Container}

\author{Ramin Golestanian}
\email{r.golestanian@sheffield.ac.uk} \affiliation{Department of
Physics and Astronomy, University of Sheffield, Sheffield S3 7RH,
UK}

\date{\today}

\begin{abstract}
A recently introduced model for an autonomous swimmer at low
Reynolds number that is comprised of three spheres connected by two
arms is considered when one of the spheres has a large radius. The
Stokes hydrodynamic flow associated with the swimming strokes and
net motion of this system can be studied analytically using the
Stokes Green's function of a point force in front of a sphere of
arbitrary radius $R$ provided by Oseen. The swimming velocity is
calculated, and shown to scale as $1/R^3$ with the radius of the
sphere.
\end{abstract}

\maketitle

Since the pioneering work of G. I. Taylor \cite{taylor} we know that
swimming at low Reynolds number is a nontrivial task, which is also
echoed in the incredibly sophisticated mechanisms involved in
swimming of bacteria \cite{berg,bray}. While nature could overcome
the difficulties of swimming at low Reynolds number by taking
advantage of a continuum of degrees of freedom, this will be much
more challenging for an artificially engineered model, as it could
often use a finite number of degrees of freedom like most manmade
devices. The relation between the number of degrees of freedom and
swimming at low Reynolds number was considered by Purcell in his
seminal work \cite{purcell1}, who concluded that one degree of
freedom is not enough and the minimum required is two. He further
elaborated on this by proposing a simple swimmer model made of three
joined rods, whose motion was fully analyzed only recently
\cite{howard1} due to the complicated nature of the hydrodynamic
problem of three rotating finite rods \cite{mit}.

If we take advantage of the positional degrees of freedom rather
than the orientational ones, we can construct swimmer models that
can be much more easily analyzed. This is the principle used in the
design of a simple swimmer that is made of three spheres connected
by two arms that open and close \cite{3SS}. The simplicity of the
three-sphere swimmer model allows us to study various characteristic
properties of the swimmer itself---such as the effects of external
load and noise, stress distribution, power consumption, etc.---in
great details \cite{Re3SS,chem3SS}. Moreover, it also makes it
possible to study the interaction between two \cite{gareth} or more
of these swimmers, which could be used towards a systematic study of
the floc behavior of a suspension of such swimmers
\cite{pedley,sriram,hern,tannie}. A similar principle has been used
to propose other swimmer models that can also be easily analyzed
\cite{avron,feld,earl}. This class of swimmer models uses a finite
number of compact degrees of freedom, and is to be contrasted from
those that use a continuum \cite{howard2,NG}. An artificial
microscopic swimmer has been recently made using magnetic beads
connected by DNA linkers, and has been shown to be able to propel
itself via beating strokes that are induced by an oscillating
magnetic field \cite{Dreyfus}. It is also possible to design
swimmers with no moving parts by taking advantage of phoretic
phenomena, as has been proposed and studied theoretically
\cite{GLA-1,GLA-2,Kapral} and also realized experimentally
\cite{chemloc1,chemloc2,Howse}.

The analysis of the three-sphere swimmer model has so far been made
in the limit where the radii of the spheres are the smallest length
scales in the system, namely, they are much smaller than the
distances between the spheres. In practical applications, however,
one can imagine that the swimmer could be used to carry a cargo,
which would be possible if one of the spheres was blown into a large
hollow spherical shell, with a radius that is comparable to the
distance between the spheres. A similar geometry could also be
useful for experimental testing of the swimmer design, as a large
spherical bead as a part of the swimmer would allow optical probing
of its propelled motion. With these motivations, here we study the
swimming of the three-sphere model with one of the spheres having a
finite radius comparable to the other length scales. It is possible
to fully analyze the motion of this model swimmer along the same
lines as the simpler version, thanks to the Stokes Green's function
of a point force in front of a sphere of arbitrary radius calculated
by Oseen in 1927 using the method of images \cite{Oseen}. We find
that the average swimming velocity scales as $1/R^3$ with the radius
of the container. Our results show that a simplistic view of
``estimating'' the velocity of an autonomous swimmer by dividing the
propelling force in the system by its Stokes friction coefficient is
incorrect, and a only proper analysis of the hydrodynamic flow due
to the deformations of the swimmer should be used to determine the
propulsion velocity.

\begin{figure}
\includegraphics[width=.75\columnwidth]{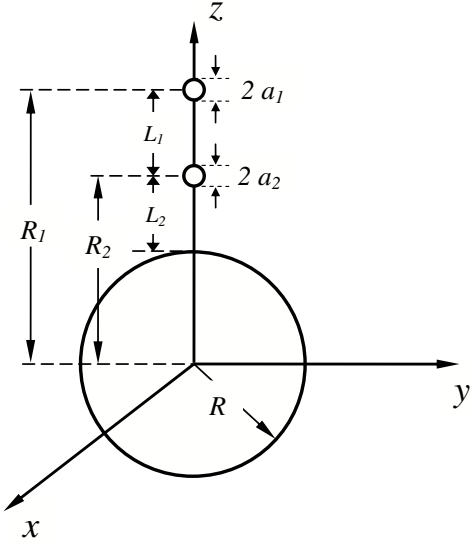}
\caption{Schematics of the swimmer with a cargo container. The
instantaneous distance between the two small spheres and the center
of the large sphere are $R_1(t)$ and $R_2(t)$, and there three
spheres are assumed to always be positioned along the $z$-axis.}
\label{fig:sphere}
\end{figure}

Consider the geometry of Fig. \ref{fig:sphere} where one of the
spheres in the swimmer is in the form of a large container of radius
$R$ and the other two have radii $a_1$ and $a_2$. To calculate the
swimming velocity of the machine, we work in the frame of reference
that is co-moving with the large sphere. Assuming that there are two
force centers located at ${\bf x}_1$ and ${\bf x}_2$ exerting the
forces ${\bf f}_1$ and ${\bf f}_2$, respectively, we can write the
velocity of the fluid at any point ${\bf x}$ as
\begin{eqnarray}
v_i({\bf x})&=&-V_i+\frac{1}{4}\frac{R}{r}
\left(3+\frac{R^2}{r^2}\right) V_i \nonumber \\
&+&\frac{3}{4}\frac{R}{r} \left(1-\frac{R^2}{r^2}\right)\frac{x_i x_j}{r^2} V_j \nonumber \\
&+&\frac{1}{8 \pi \eta} \left[G_{ij}({\bf x},{\bf x}_1)
f_{1j}+G_{ij}({\bf x},{\bf x}_2) f_{2j}\right],\label{v(x)-1}
\end{eqnarray}
where ${\bf V}$ is the velocity of the large sphere in the
laboratory frame, and $G_{ij}({\bf x},{\bf x}')$ is the Oseen
Green's function that vanishes on the surface of the large sphere
\cite{Oseen,Pozrikidis}.

Assuming that the two small spheres are located on the $z$-axis (see
Fig. \ref{fig:sphere}) at ${\bf x}_1=(0,0,R_1)$ and ${\bf
x}_2=(0,0,R_2)$, symmetry requires that the device swims along the
$z$-direction, namely, ${\bf V}=(0,0,V)$. Using no-slip boundary
condition on the two small spheres, and defining the quantities
\begin{equation}
C_\alpha=1-\frac{3}{2}\frac{R}{R_\alpha}+\frac{1}{2}\frac{R^3}{R_\alpha^3},\label{C-alpha}
\end{equation}
we can find their velocities as
\begin{equation}
v_\alpha=-C_\alpha V+\frac{1}{8 \pi \eta} \;{\cal H}_{\alpha
\beta} f_\beta,\label{v-alpha}
\end{equation}
where ${\cal H}_{\alpha \beta}=G_{zz}({\bf x}_\alpha,{\bf x}_\beta)$
for $\alpha,\beta=1,2$, and summation over repeated indices is
understood. Note that ${\cal H}_{12}={\cal H}_{21}$ due to the
symmetry of the Green's function. Assuming that the small spheres
have prescribed velocities $v_\alpha=\dot{R}_\alpha$, we can
determine the forces as
\begin{equation}
f_\alpha=8 \pi \eta \; {\cal H}^{-1}_{\alpha \beta}\;
(\dot{R}_\beta+C_\beta V),\label{f-alpha}
\end{equation}
where ${\cal H}^{-1}_{\alpha \beta}$ represent the components of the
inverse matrix for ${\cal H}_{\alpha \beta}$, with respect to the
$\alpha,\beta$ indices. Note that strictly speaking Faxen's theorem
\cite{Faxen} requires that an additional contribution proportional
to the Laplacian of the velocity field is added to the expression
for the velocity of the small spheres. However, these contributions
will generate terms that are higher order in the ratio between the
sizes of the small spheres and the distances in the system, and can
hence be neglected here.

We can now impose the condition of zero external force or the
vanishing of the total Stokeslet strength as \cite{Higdon}
\begin{equation}
C_\alpha f_\alpha+6 \pi \eta R V=0,\label{stokeslet=0}
\end{equation}
which yields the final equation. The above equation can be
understood as the vanishing of the sum of the forces exerted on the
fluid at the outer surface of the sphere. When we have extended
objects rather than just point forces, we should enforce the
force-free condition by looking at the stress field in the vicinity
of the swimmer and show that its integral over any closed surface
vanishes. In our case here, this stress will have a contribution
from the sphere, which yields the familiar Stokes force term in Eq.
(\ref{stokeslet=0}), plus contributions that have propagated from
the point forces that are located at distances $R_1$ and $R_2$. This
form of the force-free constraint can be related to the case of
three point-like spheres, where the force-free relation reads
$f_1+f_2+f_3=0$. To this end, we can simply take the limit of $R \to
0$ in Eq. (\ref{stokeslet=0}) and note that the viscous force term
will produce $f_3$ in this limit as $V$ is the velocity of the third
sphere. Equation (\ref{stokeslet=0}) proves that it is incorrect to
assume that the velocity of the bead can be extracted from a force
balance between the contributions from the two force centers and the
friction force at the bead, which would have (falsely) implied the
vanishing of $\sum_\alpha f_\alpha+6 \pi \eta R V$.

Putting in the forces from Eq. (\ref{f-alpha}), Eq.
(\ref{stokeslet=0}) can be solved to yield to propulsion velocity as
\begin{equation}
V=-\frac{C_\alpha {\cal H}^{-1}_{\alpha \beta}
\dot{R}_\beta}{C_\alpha {\cal H}^{-1}_{\alpha \beta}
C_\beta+\frac{3}{4} R}.\label{V-final-sphere}
\end{equation}
We also feed this result back into Eq. (\ref{f-alpha}) to find the
forces at the two small spheres. We find
\begin{equation}
f_\alpha=8 \pi \eta \left[ {\cal H}^{-1}_{\alpha
\beta}\dot{R}_\beta-\frac{{\cal H}^{-1}_{\alpha \beta} C_\beta
C_\gamma {\cal H}^{-1}_{\gamma \delta} \dot{R}_\delta}{C_\gamma
{\cal H}^{-1}_{\gamma \delta} C_\delta+\frac{3}{4}
R}\right],\label{f-alpha-2}
\end{equation}
where $\gamma,\delta=1,2$.

To calculate the swimming velocity we need the explicit expression
for $G_{zz}(z,z')$, which is presented in the 1927 monograph by
Oseen \cite{Oseen}. It reads
\begin{eqnarray}
G_{zz}(z,z')&=&\frac{2}{|z-z'|}-\frac{2 R}{|z z'-R^2|} \nonumber
\\
&-&\frac{R (z^2-R^2) (z'^2-R^2)}{(z z'-R^2)^3},\label{Gzz-1}
\end{eqnarray}
which yields the following expressions for the components of the
${\cal H}$-matrix
\begin{eqnarray}
{\cal H}_{11}&=&\frac{4}{3 a_1}-\frac{3 R}{(R_1^2-R^2)}, \label{H-11} \\
{\cal H}_{22}&=&\frac{4}{3 a_2}-\frac{3 R}{(R_2^2-R^2)}, \label{H-22} \\
{\cal H}_{12}&=&\frac{2}{|R_1-R_2|}-\frac{2 R}{(R_1 R_2-R^2)}
\nonumber \\
&-&\frac{R (R_1^2-R^2) (R_2^2-R^2)}{(R_1 R_2-R^2)^3}, \label{H-12}
\end{eqnarray}
where the terms involving $a_1$ and $a_2$ appear with a different
prefactor according to standard treatments of the Oseen tensor to
extract the friction coefficient. Putting the explicit forms of the
${\cal H}_{\alpha \beta}$ matrix elements as functions of the
instantaneous separations $R_1(t)$ and $R_2(t)$ in Eq.
(\ref{V-final-sphere}), we find a closed form expression for the
instantaneous swimming velocity.

It is instructive to study the asymptotic limits of Eq.
(\ref{V-final-sphere}) in various limits. In the limit where $(a_1,
a_2, R) \ll (R_1, R_2, R_1-R_2)$, we can use $R_2=\ell_2+u_2$ and
$R_1=\ell_1+\ell_2+u_1+u_2$ and expand the expression in powers of
$u_\alpha$. Keeping only terms up to the leading order, we find the
average velocity as
\begin{eqnarray}
\overline{V}&=&-\frac{3}{2} \frac{a_1 a_2 R}{(a_1+a_2+R)^2}
\left[\frac{1}{\ell_1^2}+\frac{1}{\ell_2^2}-\frac{1}{(\ell_1+\ell_2)^2}\right]\nonumber
\\
&&\times \;\overline{(u_1 \dot{u}_2-\dot{u}_1 u_2)},\label{V-bar-1}
\end{eqnarray}
where the averaging over a complete swimming cycle with period $T$
is defined by $\overline{V}=\frac{1}{T} \int_0^T d t V(t)$. This is
identical to the result obtained previously \cite{Re3SS} using the
standard form of the Oseen tensor, which is a good check for our
result. (The apparent sign difference with the previous result
\cite{Re3SS} is due to a different convention for the direction of
the $z$-axis.) Note that in this case the swimming velocity scales
as the inverse of the second power of the largest length scale in
the system, namely $1/\ell^2$.

In the limit where $(a_1,a_2) \ll (R_1-R,R_2-R) \ll R$, we can use
$R_1=R+L_1+L_2$ and $R_2=R+L_2$ (see Fig. \ref{fig:sphere}), and
expand the expression in powers of $L_\alpha/R$. We find the
instantaneous velocity as
\begin{eqnarray}
V&=&\frac{9}{2}\left(\frac{a_1 a_2}{R^3}\right)\frac{L_2^2 \;(3
L_1+4 L_2)}{L_1 (L_1+2 L_2)^3} \nonumber
\\
&&\times \left[(\dot{L}_1+\dot{L}_2) L_2^2+\dot{L}_2
(L_1+L_2)^2\right],\label{V-1/R3-1}
\end{eqnarray}
to the leading order. Expanding for small deformations using
$L_1=\ell_1+u_1$ and $L_2=\ell_2+u_2$, yields
\begin{equation}
\overline{V}=\frac{9}{4}\left(\frac{a_1
a_2}{R^3}\right)\frac{\ell_2^2 \;(3 \ell_1+2 \ell_2)}{\ell_1^2\;
(\ell_1+2 \ell_2)} \;\overline{\left(\dot{u}_1 u_2-\dot{u}_2
u_1\right)}.\label{V-1/R3-3}
\end{equation}
Equations (\ref{V-1/R3-1})  and (\ref{V-1/R3-3}) are proportional to
$1/R^3$, which is in marked contrast to the $1/\ell^2$ scaling in
the case of a (more) symmetric swimmer. The difference in the
scaling suggests that it might be more efficient to distribute the
size of the swimmer across the three spheres, rather than
concentrate all of it on one sphere. Such a statement, however,
needs to be checked in the limit of a symmetric swimmer with spheres
of large radii compared to their surface-to-surface separations.
Equation (\ref{V-1/R3-3}) also shows that it is beneficial for the
two small spheres to be close to each other and relatively farther
away from the surface of the sphere, i.e. $\ell_2 \gg \ell_1$. Note,
however, that this result [Eq. (\ref{V-1/R3-3})] is only valid for
$(a_1,a_2) \ll (\ell_1,\ell_2) \ll R$, which means that the previous
prescription for creating optimal swimming efficiency should be
written as $(a_1,a_2) \ll \ell_1 \ll \ell_2 \ll R$ rigorously
speaking.

The present calculation can be readily generalized to the case of
one large sphere with many small spheres. For a line-up of all the
spheres along the $z$-axis, Eqs. (\ref{C-alpha}), (\ref{v-alpha}),
(\ref{f-alpha}), (\ref{stokeslet=0}), (\ref{V-final-sphere}), and
(\ref{f-alpha-2}) will hold true provided the indices are run over
the number of small spheres in the system. For more complicated
assortments of small spheres, one can still use this formalism
provided the right tensorial structure is maintained along the way.
A useful particular case could be a line-up of small spheres that
oscillate laterally, i.e. perpendicular to the $z$-axis, which would
mimic the beating motion of a (finite) tail. It will also be
interesting to consider the possibility of a number of such
structures attached to the bead interacting with each other
hydrodynamically. These ``active tails'' could potentially
synchronize with each other under certain conditions, to generate an
efficient propulsion. Our systematic approach might be able to
provide useful insight into the intriguing phenomenon of collective
hydrodynamic patterns and phases \cite{sync1,sync2,sync3}. One can
also study the efficiency of the swimmer in terms of power
consumption and useful work, which for this swimmer follows very
closely similar studies that are presented elsewhere \cite{Re3SS}.

In conclusion, we have generalized the simple model of a low
Reynolds number swimmer to the case where one of the spheres is
large, and analyzed its motion using the Oseen tensor of a point
force in front of a sphere which imposes no-slip boundary condition
on the Stokes flow. The results found here could help us in possible
experimental probes of self-propelled devices.


I would like to acknowledge fruitful discussions with A. Ajdari and
A Najafi.

\end{document}